\documentstyle[aps,twocolumn,prl]{revtex}
\begin{document}
\draft
\twocolumn[\hsize\textwidth\columnwidth\hsize\csname
@twocolumnfalse\endcsname

\title{A projection method for statics and dynamics 
             of lattice spin systems}

\author{M. Kolesik,$^{1,2}$ 
        M. A. Novotny,$^{1}$ and 
        P. A. Rikvold$^{1,3}$}

\address{
$^1$Supercomputer Computations Research Institute, 
    Florida State University, 
    Tallahassee, Florida 32306-4130 \\
$^2$Institute of Physics, 
    Slovak Academy of Sciences, D\' ubravsk\' a cesta 9, 
    84228 Bratislava, Slovak Republic \\
$^3$Center for Materials Research and Technology, and 
    Department of Physics, 
    Florida State University, 
    Tallahassee, Florida 32306-4350
        }

\date{\today}
\maketitle
\begin{abstract}
A method based on Monte Carlo sampling of the
probability flows projected onto the subspace of one
or more slow variables is proposed for investigation
of dynamic and static properties of lattice spin
systems.  We illustrate the method by applying it,
with projection onto the order-parameter subspace, to
the three-dimensional 3-state Potts model in
equilibrium and to metastable decay in a
three-dimensional 3-state kinetic Potts model. 
\end{abstract}

\pacs{PACS Number(s):
      75.40.Mg, 
      05.50.+q, 
      02.70.Lq, 
      02.50.-r} 
\vskip1pc]

The widespread use of computer simulations in many
fields of physics presents a constant challenge to
develop new, faster, and more efficient algorithms.
Here we describe a method to study dynamic and
static properties of spin lattice systems. It is based
on Monte Carlo sampling of the probability flow
projected onto the subspace of one or more slow
variables. Here we use the order parameter.  The
projected information is subsequently used to
reconstruct dynamic and/or static quantities.  This
idea was first explored by Schulman \cite{Schulman},
and a similar approach for the dynamics of metastable
decay was developed by Lee {\it et al.}\cite{Lee} and
in Ref.~\onlinecite{Kolesik97}.  The present work 
extends these developments.  We show how
appropriately sampled projected probability flows can
be utilized to investigate static probability
distributions, as well as the dynamics of 
spin systems. The method is illustrated for three cases.
The first two deal with three-dimensional, three-state
Potts models in equilibrium. The ferromagnetic model
at its transition temperature is considered in order
to explain, for a generic situation, the basics of the
method. This model is of interest also in
lattice-gauge theory \cite{Svetitsky}.  Next, the
antiferromagnetic model below its critical temperature
is included to show an unusual example of the phase
structure. Although only intended as an illustration,
these results greatly elucidate previous
Monte Carlo observations of a medium-temperature phase
in this model \cite{Kolesik95,Swendsen-Wang}. Our third
example concerns the application to the dynamics of
metastable decay in the 3-dimensional 3-state
kinetic ferromagnetic Potts model, which can be
regarded as an extension of the Ising model
approximation for extremely anisotropic magnetic
systems \cite{Rikvold,Richards}. We demonstrate the
strength of the method by measuring metastable
lifetimes in a region of weak 
fields, where
direct
simulations are not feasible.

Consider a 3-state Potts model with the Hamiltonian
$
{\cal H} = 
-J \sum_{\langle i,j \rangle} \delta (\sigma_i,\sigma_j)
$
where $\sigma_i \in \{0,1,2\}$ is the ``spin'' at
lattice site $i$, and the summation runs over all
nearest-neighbor pairs on a simple-cubic lattice.
Macroscopically, the system is characterized by the
concentrations $\{n_0,n_1,n_2\}, \sum n_i$$=$$1$, of
spins in the three states.  These triples are
mapped into an equilateral triangle representing the
order-parameter space. In Figs.~1 and 2, the
projection of a point onto the axis extending from the
$i$-th corner gives the concentration $n_i$.

Consider a simulation using a Monte Carlo method with
local updates at randomly chosen sites.  At any given
moment, the spins can be divided into classes
specified by the state $\sigma$ of the spin and by the
numbers $\{a,b,6-a-b\}$ of its neighbors in the states
$\{0,1,2\}$.  Let $c^{\sigma}_{a b}(n_0,n_1,n_2)$ be
the average {\em equilibrium} population of spins in the
class $\{\sigma,a,b\}$, conditional on the total
concentrations $n_i$ of spins in state $i$. Further,
denote by $p^{\sigma \sigma'}_{a b}$ the probability
that a spin in the class $\{\sigma,a,b\}$ will flip to
the state $\sigma'$ when visited by the updating
algorithm.

The central objects of the proposed method are the
global flip rates $ v_{\sigma \sigma'} (n_0,n_1,n_2)$: 
\begin{equation}
v_{\sigma \sigma'} (n_0,n_1,n_2) = 
 \sum_{a b} c^{\sigma}_{a b}(n_0,n_1,n_2)\
 p^{\sigma \sigma'}_{a b}\ .
\end{equation}
They define an ideal aggregate Markov process\cite{Bucholtz} 
on the order-parameter space $\{(n_0,n_1,n_2)\}$. 
This process gives an {\em exact} equilibrium aggregated 
probability density $P(n_0,n_1,n_2)$\cite{Bucholtz}, 
which can be calculated from the detailed-balance condition
$
 P(n_0+1,n_1,n_2) v_{0 1}(n_0+1,n_1,n_2) = 
 P(n_0,n_1+1,n_2) v_{1 0}(n_0,n_1+1,n_2)
$
plus analogous equations for the other transitions \cite{unpub}.

Of special importance are the zeroes of the ``drift
functions'' $v_{\sigma \sigma'} - v_{\sigma' \sigma}$,
since they determine the extremal points of the
probability densities. For the statics of the Potts
model, we restrict ourselves to these zero loci.

\begin{figure}[t]
\vspace*{2.6in} 
\includegraphics{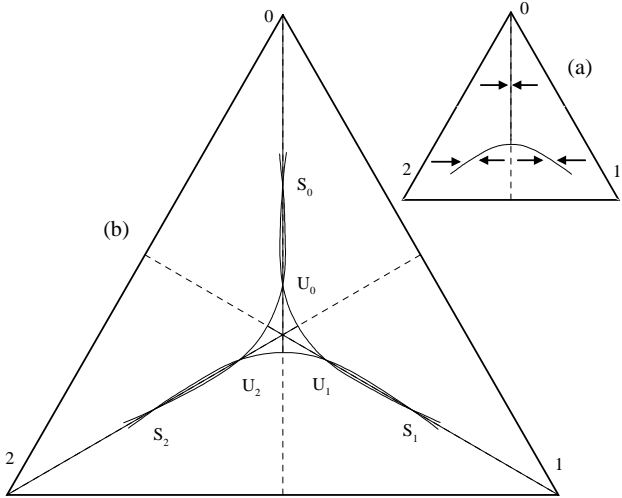} 
\caption{ 
(a): Zeros of the drift $v_{1 2}-v_{2 1}$ in the
$1\leftrightarrow 2$ direction for the 3-dimensional
3-state Potts ferromagnet.  Light solid lines are
stable loci, while dashed lines are unstable.  Heavy
arrows indicate the direction of the probability flow.
(b): Zero loci for the three directions superimposed.
See text for a full description.
\label{fig:FPotts}
}
\end{figure}

We simulated the ferromagnetic model ($J$$=$$1$) at its
first-order phase transition temperature $T$$=$$1.81618$
\cite{Wilson} for a $20^3$ lattice. 
Flip-rate histograms were binned using
$2^{14}$ equilateral triangles covering the whole
order-parameter triangle, and $10^6$ configurations
were typically generated.  Figure \ref{fig:FPotts}(a)
shows a schematic picture of the drift in one of the
three directions. Its zero locus consists of two
parts. The first one is the symmetry axis of the
triangle, and the second is a symmetric arc.
We only show a
portion of the arc, since the statistics become
insufficient far from the probability density maxima. 
Flows in different directions are related by rotations
of $2\pi/3$.  All three zero loci are superimposed in
Fig.~\ref{fig:FPotts}(b).  Where three loci
corresponding to different directions intersect, the
probability density has a stationary point.  There are
three stable extrema $S_i$, corresponding to the three
ordered phases. The three points $U_i$ are saddle
points, and the center of the triangle represents the
stable, disordered phase. This example shows how one
identifies candidates for stable phases from the
intersections of the drift-function zero loci.

\begin{figure}[t]
\vspace*{2.0in} 
\includegraphics{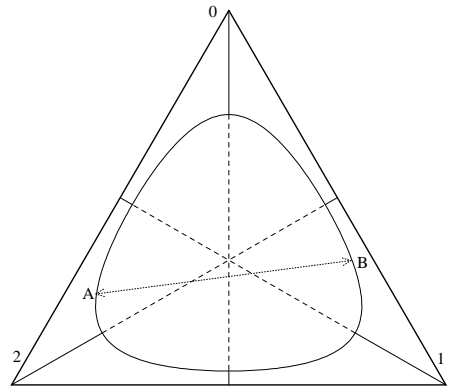} 
\caption{  
Zero locus of the drift functions $v_{\sigma
\sigma'}-v_{\sigma' \sigma}$ of the 3-dimensional
3-state Potts antiferromagnet at $T=1.0J$. The dashed
sections of the triangle axes are unstable in
directions perpendicular to the axes. The solid parts
are stable. The closed, solid curve is stable in all
three directions.  This curve, obtained for a $64^3$
lattice, represents the degenerate maximum of the
probability density of a sublattice location in the
order-parameter space. 
\label{fig:APotts}
}
\end{figure}

Next, we turn to the antiferromagnetic Potts model
($J$$=$$-1$).  Because of the sublattice-symmetry
breaking, we sampled the flip rates for each
sublattice separately.
Figure \ref{fig:APotts} shows the union of the zero
loci of the drift functions in all three directions.
As in the ferromagnetic case, for each direction there
is the symmetry induced straight line (triangle axis)
and a nontrivial part, which is here a closed curve. 
The closed-curve
parts of the zero loci have several interesting
properties: 

1. The closed loci for the three different directions are 
{\em identical} and can be accurately parameterized by
\begin{equation}
\{ m \cos t + r \cos 2t , m \sin t - r \sin 2t  \} \ , 
      t\in \langle 0, 2\pi \rangle .
\label{eq:param}
\end{equation}

2. Positions of the sublattices on this curve are
correlated in such a way that their distance
(sublattice magnetization difference $| AB |=2m$ in
Fig. \ref{fig:APotts}) is constant. 

3. There are finite-size effects in the diameter and
shape of the curve, but there is no sign that it
separates into three distinct components.

We stress that the only observed deviations from these
properties are numerical uncertainties on the order of
the discreteness of the order-parameter space. On
lattices smaller than $32^3$, we sampled complete
flip-rate histograms as in the ferromagnetic case. On
larger lattices, we only measured flip rates at
isolated points of the order-parameter space
(typically $5\times 10^4$ configurations).
The data were then interpolated and used to find
the roots of the drift functions.  In this way, we
could confirm the above observations, even with
lattices as large as $64^3$, by measuring flip rates
only in the vicinity of the expected zero locus. In
fact, having located two points we can predict the
rest with an accuracy suggesting that the
parameterization of Eq.~(\ref{eq:param})  may be exact.

The above observations mean that the probability
density has a degenerate maximum. Whereas the
magnitude of the sublattice magnetization difference
is fixed, it can point in an arbitrary direction.
Thus, a system with only three microscopic states
exhibits a continuous symmetry on the macroscopic
scale.  This is the rotationally symmetric phase which
was found in recent Monte Carlo simulations
\cite{Kolesik95,Swendsen-Wang}. However, the
conventional probability density sampling technique
did not provide completely convincing evidence that
this symmetry is not (weakly) broken.  The present
approach gives much stronger support for the
symmetric phase. It locates the extrema of the
probability density without sampling the distribution
itself.  With relatively modest statistics, the
symmetric property was corroborated with an accuracy
approaching the limit imposed by the discreteness of
the order-parameter space. Such precision would
require an enormous effort if one were to
investigate the sampled distribution directly.

Our final example deals with metastable decay in the
ferromagnetic Potts model with the Hamiltonian 
$
{\cal H} = 
-J \sum_{\langle i,j \rangle} \delta (\sigma_i, \sigma_j)
+ H \sum_i [ \delta (0, \sigma_j) - \delta (1, \sigma_j)].  
$ 
Here, we have added a term describing the interaction
with the external field $H$.  With this choice of the
external field, the model can be regarded as an
extension of the kinetic Ising model, which can serve
as an approximation for nanoscale ferromagnets
\cite{Rikvold,Richards}. The third spin state of the
present model can mimic local magnetization
``perpendicular'' to the external field and allows for
``finite anisotropy.'' The Glauber dynamics with
updates at randomly chosen sites is used throughout
the rest of the paper. Time is measured in Monte Carlo
Steps per Spin (MCSS). 

An essential quantity related to metastability is
the lifetime $\tau$, defined as follows. All spins are
initialized in state $0$, the temperature $T<T_c$ is fixed,
and an external
magnetic field $H$ favoring state $1$ and disfavoring
state $0$ is applied. The field does not interact with
spins in state $2$.  This initial state is metastable
and decays through the nucleation and subsequent
growth of stable-phase droplets. The average time
needed to reach a configuration with half of the
system in the stable phase is $\tau$. The
difficulty is that realistic models, subjected to
``experimentally reasonable'' magnetic fields, have
lifetimes which are extremely long in terms of the
Monte Carlo time. Here we describe a significant
extension of the method proposed in
Refs.~\onlinecite{Schulman,Lee,Kolesik97}. This allows
us to obtain lifetimes in arbitrarily weak fields
without prohibitively lengthy simulations.

The projected flip rates are defined, as in the static
case, in terms of the normalized spin-class populations
$c^{\sigma}_{ab}(n)$ and flipping probabilities
$p^{\sigma \sigma'}_{ab}$: 
\begin{equation}
g(n)=\sum_{ab,\sigma\ne 1} c^{\sigma}_{ab}(n)\  p^{\sigma 1}_{ab}\ , \
s(n)=\sum_{ab,\sigma\ne 1} c^{1}_{ab}(n)\  p^{1 \sigma}_{ab} .
\label{eq:gsrates}
\end{equation}
We parameterize $g$ and $s$ only by the total
number $n$ of spins in state $1$, thus projecting  
the population data onto a one-dimensional histogram.  
The rates $g$ and $s$ correspond to growth and shrinkage 
of the stable phase. They depend on the external field 
and on the way the configurations are generated, as 
explained below.

The flip rates are used to map the metastable-decay
dynamics onto a one-dimensional absorbing Markov
chain. We assign to all configurations with $n$
overturned spins a single state $n$ in the chain. The
one-dimensional dynamics is given by the flip rates.
{}From state $n$ we have the probability $g(n)$ of
jumping to state $n+1$, the probability $s(n)$ of
jumping to $n-1$, and the probability $1- s(n)- g(n)$
of remaining in the current state.  This random walk
starts at $n=0$ and terminates when it reaches $n=N$,
corresponding to a stopping criterion which we chose
to be $N = V/2$, with $V$ the volume of the system. 
Using standard methods from the theory of absorbing
Markov chains \cite{Lee,Novotny}, we obtain the mean
lifetime $\tau$ and the total average time $h(n)$
(measured in MCSS, with $h(N)=0$) spent by the random
walker in the state $n$ as
\begin{equation} 
 \tau = \sum_{n=0}^{N-1} h(n)\ \ , \ \
 h(n-1) = { V^{-1} + s(n) h(n) \over g(n-1)} \ .  
\label{eq:lifetime}
\end{equation} 
How accurately these formulas reproduce the lifetime
depends on how the class populations are
sampled.  One option is to sample them
in zero external field in an equilibrium ensemble with
conserved order parameter for each needed value of $n$
\cite{Schulman,Lee}. Such data can be used to estimate
the lifetime in very weak fields, but in strong fields 
it is underestimated
because the class populations near the
top of the free-energy barrier are not reproduced well
by the equilibrium ensemble.

The simple but important improvement presented here is
the way the class populations are measured. At any
time, the system is only allowed to have $n$, the
number of spins in the stable phase, larger than a
time-dependent lower bound, $n_{\rm min}$. Simulation
starts with $n_{\rm min}(t$$=$$0)=-1$, and $n_{\rm min}$ is
increased slowly.  The class populations
$c_{ab}^{\sigma}$ are sampled during this
forced-escape simulation with an applied external
field and are subsequently used to calculate the
lifetime from Eqs.~(\ref{eq:gsrates},\ref{eq:lifetime}). 
Why does this work?  
Consider first the limit of zero forcing speed  
($n_{\rm min}$$=$$-1$ at all times).
One starts a simulation by releasing a ``random walker''
from the initial state. 
When the walker reaches the absorbing state,
one starts a new run 
and repeats the whole process until $N_{\rm esc}$ escapes
from metastability are realized. One can imagine that each
walker generates an oriented ``world line.'' Therefore, 
an ``equation of continuity'' holds for each $n<N$:
$
N_{n\to n+1} = N_{\rm esc} + N_{n+1 \to n} 
$
with $N_{i\to j}$ the number of transitions between the
subspaces with $i$ and $j$ overturned spins. 
It is straightforward to express this equality in terms of the
transition probabilities and class populations as generated
by the simulation. One obtains a relation for 
$h(n)$ equivalent to Eq.~(\ref{eq:lifetime}) \cite{unpub}.
Thus, if the forcing is infinitely slow, 
Eqs.~(\ref{eq:gsrates},\ref{eq:lifetime}) give the exact $\tau$.

Now, if the rate of increase of $n_{\rm min}$
is nonzero but sufficiently small, the system always 
produces nearly the
``correct'' configurations as if there were no
forcing. While deep in the metastable free-energy
well, forcing prevents the system from returning to
$n\le n_{\rm min}$, but it still allows it to thermalize
and generate metastable configurations with $n$ at and
slightly above $n_{\rm min}$. As $n_{\rm min}$
increases, the procedure scans the configurations
along the escape path from metastability.  When
$n_{\rm min}$ approaches the top of the free-energy
barrier, the system has a better chance to escape, and
the stable phase grows too quickly to equilibrate. 
There is nothing to prevent escape, because there is
no upper bound on $n$ which would  cause
unwanted thermalization.  The system is free to
escape through  natural nonequilibrium 
configurations.

To approach the slow-forcing limit in practice, 
one performs a series
of measurements and determines a value of $d n_{\rm
min}/dt$ below which the estimated lifetime is
insensitive to the choice of $d n_{\rm min}/dt$. 
We expect such forcing speed to be related to the 
inverse equilibration time for states deep in the 
metastable well. 
Another computationally important aspect is to obtain
sufficient statistics for the forced escapes, since
the class-population data tend to be noisy at and
beyond the top of the free-energy barrier. Here, we measured
about $10^3$ escapes at rates up to $d n_{\rm min}/dt
\approx 10^{-3}$ MCSS$^{-1}$.

\begin{figure}[t]
\vspace*{2.3in} 
\includegraphics{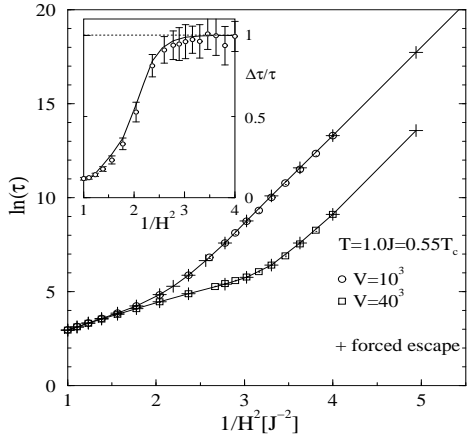} 
\caption{
The metastable lifetime of a 3-dimensional 3-state
kinetic Potts ferromagnet as a function of the
magnetic field for two system sizes at 
$T\approx 0.55T_c$. Symbols $\circ$ and $\Box$ are
simulation results (with error bars smaller than the
symbol size), and lines connect the points ($+$) 
calculated from the class-population data sampled by
the forced-escape method.  The inset shows a
comparison of the measured (circles) and calculated
(line) relative standard deviation of the lifetime. 
\label{fig:Potts_tau}
}
\end{figure}

Figure~\ref{fig:Potts_tau} shows a comparison between
the lifetimes obtained from direct simulations and
those calculated with the forced-escape method. The
agreement is excellent for lifetimes in the whole
region accessible to direct simulations. Moreover, the
forced escape method can provide lifetime estimates
deep in the region of weak fields, where direct
simulation is practically impossible. We emphasize
that having measured the class populations, one can
utilize them to calculate lifetimes for an arbitrary
local dynamic with random updates. We can e.g.\ calculate 
what the lifetimes would be if we used the
Metropolis instead of the Glauber dynamic simply by
replacing the flip probabilities $p_{a b}^{\sigma
\sigma'}$.  
Although different dynamics produce different flip
rates, the class-populations remain close to local
equilibrium and are therefore similar for all dynamics
that obey detailed balance.

In a similar way as the average lifetime $\tau$, one can use
the flip rates to calculate higher moments of the
lifetime probability distribution \cite{Lee}. The
inset in Fig.~\ref{fig:Potts_tau} shows the relative
standard deviation $\Delta\tau/\tau$ as a function of
the field. Despite the fact that the higher moments, 
unlike the mean lifetime,
are only approximate even in the slow-forcing limit,
the agreement between the directly
simulated results and those based on the forced-escape
method (solid line) is good. We believe this is because
the relevant Markov chains are nearly weakly 
lumpable\cite{Bucholtz,weaklump} with respect the states along 
the escape from metastability. For a quantitative
description of the form of $\tau$ and $\Delta\tau /\tau$,
see Refs.~\onlinecite{Rikvold} and
\onlinecite{Richards}.

In summary, we propose a new method to study the
dynamic and static properties of lattice spin systems.
It consists in Monte Carlo sampling of the spin-class
populations in a projected subspace. These are
subsequently used to calculate the spin-flip rates. In
the static case, the knowledge of the flip rates is
equivalent to the information contained in the
probability density. However, the accuracy of the
method is much better than with direct distribution
sampling.  We have demonstrated this on the example of
the 3-state antiferromagnetic Potts model.

The flip rates can be also utilized to obtain dynamic
characteristics of the systems, such as lifetimes, in
very weak fields where ordinary simulations are not
feasible. Although the resulting dynamic is only an
approximation, it provides accurate estimates. The
detailed information about the class populations
enables one to calculate the lifetime for an arbitrary
dynamic with updates at randomly chosen sites,
independently of the dynamic used during the data
sampling.

We appreciate helpful discussions with S. W. Sides.
This research was supported by 
NSF Grants No. DMR-9520325 and DMR-9634873,
FSU-MARTECH and FSU-SCRI 
(DOE Contract No. DE-FC05-85ER25000).

\end{document}